\documentclass[aps,prl,twocolumn,amsfonts,amssymb,amsmath,showpacs
nofootinbib,groupedaddress,superscriptaddress,citesort]{revtex4}

\usepackage[dvips]{graphicx}

\begin{document}
\title{Evolution equation of entanglement for general bipartite systems}
\vspace{2ex}

\author{Zong-Guo Li}
\affiliation{Beijing National Laboratory for Condensed Matter
Physics, Institute of Physics, Chinese Academy of Sciences, Beijing
100080, China}
\author{Shao-Ming Fei}
 \affiliation{Department of Mathematics, Capital Normal University, Beijing
 100037, China}
 \affiliation{Institut f{\"u}r Angewandte Mathematik, Universit{\"a}t Bonn,
53115, Germany}
\author{Z. D. Wang}
\affiliation{Department of Physics and Center of Theoretical and
Computational Physics, University of Hong Kong, Pokfulam Road, Hong
Kong, China}
\author{W. M. Liu}
\affiliation{Beijing National Laboratory for Condensed Matter
Physics, Institute of Physics, Chinese Academy of Sciences, Beijing
100080, China}

%\date{\today}

\begin{abstract}
We explore how entanglement of a general bipartite system evolves
when one subsystem undergoes the action of an arbitrary noisy
channel. It is found that the dynamics of entanglement for general
bipartite systems under the influence of such channel is determined
by the channel's action on the maximally entangled state, which
includes as a special case the results for two-qubit systems [Nature
Physics 4, 99 (2008)]. In particular, for multi-qubit or qubit-qudit
systems, we get a general factorization law for evolution equation
of entanglement with one qubit being subject to a noisy channel. Our
results can help the experimental characterization of entanglement
dynamics.
\end{abstract}


\maketitle

\emph{Introduction}---In quantum information theory, entanglement is
a vital resource for some practical applications such as quantum
cryptography, quantum teleportation and quantum computation
\cite{bennett,nielsen}. To fulfill such tasks by constructing
suitable quantum devices, we inevitably encounter some interactions
of the multiparticle quantum states under consideration with its
environment. These undesired couplings give rise to decoherence,
which degrades the entanglement when the particles propagate or the
computation evolves. Therefore, it is of great practical importance
to investigate the dynamics of entanglement for the quantum systems
under the influence of decoherence.

Recently much effort has been devoted to understanding the dynamics
of entanglement \cite{thomas,yuting,dodd,karol,markus,frank}. In
stead of deducing the evolution of entanglement from the time
evolution of the state, Thomas Konrad \emph{et al}. \cite{thomas}
provided a direct relationship between the initial and final
entanglement of an arbitrary bipartite state of two qubits with one
qubit subject to incoherent dynamics, where qubit represents the
state of 2-dimensional quantum system. It is also discussed in
\cite{markus} for two-qudit systems with either system undergoing an
arbitrary physical process, where qudit denotes the state of
D-dimensional quantum system. On the condition that the pure initial
state has $d$ non-zero Schmidt coefficients, an evolution equality
is satisfied during an initially finite time interval.

In fact, for practical applications in quantum information
processing, multipartite entanglement are often concerned, e.g.
cluster states used as a resource for one-way quantum computing
\cite{cluster}, multi-photon entangled states \cite{Pan} etc. In
this Letter, we investigate the evolution of entanglement for
multipartite pure states with one part of the system undergoing the
action of an arbitrary noisy channel, which represents the influence
of an environment, of measurements or of both. Basically this is
equivalent to studying the evolution of entanglement for bipartite
systems with one subsystem subject to some arbitrary noisy channels.
Moreover, bipartite systems with higher-dimension can improve the
performance of various quantum information and computation tasks,
such as quantum cryptography \cite{crypt}. Thus it is necessary to
investigate the dynamics of entanglement for an $N_1 \times N_2$
bipartite system under the influence of decoherence. In the
following, we find that the dynamics of entanglement for general
bipartite systems under the influence of a noisy channel is
determined by the channel's action on the maximally entangled state,
instead of exploring the time-dependent action of the channel on all
initial states. Therefore the robustness of entanglement-based
quantum information processing protocols is easily and fully
characterized by a single quantity. As applications we discuss two
examples in detail: the entanglement evolution for a generalized
three-qubit W state \cite{w} with one qubit undergoing the action of
phase noise channel and generalized amplitude damping channel
respectively, and the one for the ground state in a nuclear magnetic
resonance (NMR) system with one subsystem subject to a decay
channel.

\emph{Evolution equation of entanglement}---We first define the
entanglement measure for bipartite systems.
 For a pure state
$|\chi\rangle=\sum^{N_1}_{i=1}\sum^{N_2}_{j=1}A_{ij}|ij\rangle
\in\mathbb{C}^{N_1}\otimes\mathbb{C}^{N_2}$ in the computational
basis $|i\rangle$ and $|j\rangle$ of Hilbert space
$\mathbb{C}^{N_1}$ and $\mathbb{C}^{N_2}$ respectively, we define
the concurrence matrix $\textbf{C}$ with entries
$C_{\alpha\beta}=\langle\chi|(L_\alpha\otimes
L_\beta)|\chi^{*}\rangle$, where $|\chi^{*}\rangle$ is the complex
conjugate of $|\chi\rangle$, $L_\alpha,$
$\alpha=1,\cdots,N_1(N_1-1)/2$, and $L_\beta,$ $\beta=1,\cdots,
N_2(N_2-1)/2$, are the generators of $SO(N_1)$ and $SO(N_2)$ groups
respectively. The Frobenius norm of $\textbf{C}$ is just the
I-concurrence \cite{rungta},
\begin{eqnarray}
\label{vec1}
C[|\chi\rangle]\!=\!||\textbf{C}[|\chi\rangle]||_F\!=\!\sqrt{\sum_{\alpha=1}^{N_1(N_1-1)/2}
\sum_{\beta=1}^{N_2(N_2-1)/2}|C_{\alpha\beta}|^2},
\end{eqnarray}which reduces to concurrence when
restricting to $2\otimes 2$ systems \cite{wootters}. These
quantities can be measured for pure states \cite{measureable}. As we
konw, the I-concurrence is equal to the length of the concurrence
vector \cite{wootters2,v1}.

The $N_1(N_1\!-\!1)/2$ generators $L_{\alpha}$ of $SO(N_1)$ have the
following form, $ L_{\alpha_{(kl)}}\!=\!(-1)^{k+l+1}|k\rangle\langle
l|\!+\!(-1)^{k+l}|l\rangle\langle k|,~k\!<\!l, $ where
$\alpha_{(kl)}\!=\!(j_1,j_2,\cdots, j_{N_1-2})$ with $1\!\leq\!
j_1\!<\!j_2\!<\!\cdots\!<\! j_{N_1-2}\!\leq\! N_1$, $k\!<\!l$ and
$k,l\not\in\{j_1,j_2,\cdots, j_{N_1-2}\}$. Therefore, one has $
C_{\alpha_{kl}\beta_{k'l'}}
\!=\!2(-1)^{k+l+k'+l'}(A_{kk'}A_{ll'}-A_{kl'}A_{lk'})^{*} $ and $
C[|\chi\rangle]\!=\!2\sqrt{\sum_{i<j}^{N_1}\sum_{k<l}^{N_2}|A_{ik}A_{jl}-A_{il}A_{jk}|^2},
$ which is the just the generalized concurrence in \cite{albev} up
to a constant factor. $C[|\chi\rangle]$ is zero when $|\chi\rangle$
is separable, i.e., $A_{ij}\!=\!b_ic_j$ for some complex numbers
$b_i$, $c_j$. On the other hand, $C$ takes its maximum value
$\sqrt{2(N-1)/N}$ with $N\!=\!\mathrm{min}(N_1,N_2)$, when
$|\chi\rangle$ is a maximally entangled state.

For a bipartite mixed state $
\rho\!=\!\sum_ip_i|\psi_i\rangle\langle\psi_i|$,
 $p_i\!\geq\!0$, $\sum_ip_i\!=\!1,$ the concurrence is defined
by the convex-roof:
\begin{eqnarray}
\label{vec0} C[\rho]=\mathrm{min}\sum_ip_iC[|\psi_i\rangle],
\end{eqnarray}
where  $C[|\psi_i\rangle]$ is the norm of concurrence matrix
$\textbf{C}[|\psi_i\rangle]$ and the minimum is obtained over all
possible pure state decompositions $|\psi_i\rangle$ of $\rho$.

Let $|\chi\rangle$ be the bipartite pure initial state, and let the
second subsystem undergo the action of a noisy channel. We will
denote the noisy channel by \$ thereafter. Then the final state of
the system takes the form $\rho'\!=\!(
I\otimes\$)|\chi\rangle\langle\chi|. $ To investigate the properties
of the entanglement of the final state $\rho'$, it is convenient to
re-express the initial state as: $|\chi\rangle\!=\!(M_{\chi}\otimes
I)|\phi\rangle, $ where
$M_{\chi}\!=\!\sqrt{N_2}\sum_{i=1}^{N_1}\sum_{j=1}^{N_2}A_{ij}|i\rangle
\langle j|$ is the filtering operation \cite{gisin} acting on the
first subsystem of maximally entangled pure state,
$|\phi\rangle\!=\!\sum_{n=1}^{N_2}|n\rangle\otimes|n\rangle/\sqrt{N_2}$,
and $ I$ is the $N_2\times N_2$ identity matrix. Due to the fact
that $M_{\chi}$ and \$ act on the first and second subsystem of the
state $|\phi\rangle$ respectively, the evolution of $|\chi\rangle$
takes the form: $\rho'\!=\! (M_{\chi}\otimes
I)\rho_{\$}(M_{\chi}^{\dag}\otimes I) $, where $\rho_{\$}\!=\!(
I\otimes\$)|\phi\rangle\langle\phi|$.

In terms of concurrence matrix, we have the following theorem: if
$\rho_{\$}$ is a pure state, the concurrence $C[\rho']$ for the
state $\rho'$ is given by
\begin{eqnarray}
\label{ev4}
C[\rho']\!=\!\sqrt{\sum_{\alpha=1}^{N_1(N_1-1)/2}\quad\sum_{\beta=1}^{N_2(N_2-1)/2}|C'_{\alpha\beta}|^2},
\end{eqnarray}
where
$
C'_{\alpha\beta}=\frac{N_2}{2}\sum_{\gamma=1}^{N_2(N_2-1)/2}C_{\alpha
\gamma}[|\chi\rangle]C_{\gamma\beta}[\rho_{\$}].
$
$C_{\alpha \gamma}[|\chi\rangle]$ and $C_{\gamma\beta}[\rho_{\$}]$
are the entries of the concurrence matrices
$\textbf{C}(|\chi\rangle)$ and $\textbf{C}(\rho_{\$})$ respectively.

Let us now prove this theorem. Suppose the pure state $\rho_{\$}$
has the following generic form, $
\rho_{\$}\!=\!|\psi\rangle\langle\psi|$, $
|\psi\rangle\!=\!\sum^{N_2}_{i,j=1}a_{ij}|ij\rangle. $ The final
state has the form $\rho'\!=\!|\psi'\rangle\langle\psi'|$, where
$|\psi'\rangle\!=\!\sum^{N_1}_{i=1}\sum^{N_2}_{j=1}B_{ij}|ij\rangle$,
and $ B_{ij}\!=\!\sqrt{N_2}\sum_{l=1}^{N_2}A_{il}a_{lj}$. Some
straightforward algebra yields the equation
\begin{eqnarray}
C'_{\alpha_{kl}\beta_{k'l'}}&\!=\!&2(-1)^{k+l+k'+l'}(B_{kk'}B_{ll'}-B_{kl'}B_{lk'})^{*}\nonumber\\
%&=&2N_2\sum_{1\leq i<i'\leq N_2}(-1)^{k+l}(A_{ki}A_{li'}-A_{ki'}A_{li})^{*}\nonumber\\
%&&\times(-1)^{k'+l'}(a_{ik'}a_{i'l'}-a_{il'}a_{i'k'})^{*}\nonumber\\
%&=&\frac{N_2}{2}\sum_{1\leq i<i'\leq N_2}\Big(C_{\alpha_{kl}\gamma_{ii'}}[|\chi\rangle]C_{\gamma_{ii'}\beta_{k'l'}}[\rho_{\$}]\Big)\nonumber\\
&\!=\!&\frac{N_2}{2}\sum_{\gamma=1}^{N_2(N_2-1)/2}\Big(C_{\alpha_{kl}
\gamma}[|\chi\rangle]C_{\gamma\beta_{k'l'}}[\rho_{\$}]\Big).
\end{eqnarray}
Hence the concurrence takes the form (\ref{ev4}).
\hspace{50pt}$\blacksquare$

Remark: With respect to the relations
${C}[\rho']={C}[|\chi\rangle]{C}[\rho_{\$}]$ in \cite{thomas} for
two-qubit systems ($N_1=N_2=2$), here we have a similar relation for
the corresponding concurrence matrices,
$\textbf{C}[\rho']=\frac{N_2}{2}\textbf{C}[|\chi\rangle]\textbf{C}[\rho_{\$}]$.

For a general channel $\$$, the state $\rho_{\$}$ is usually a mixed
one. Assume $\rho_{\$}$ has an optimal pure state decomposition
$\rho_{\$}\!=\!\sum_ip_i|\phi_i\rangle\langle\phi_i|$ such that
$C[\rho_{\$}]\!=\!\sum_ip_iC[|\phi_i\rangle].$ By convexity we have
$C\big[( I\otimes\$)|\chi\rangle\langle\chi|\big]
\!=\!C\big[\sum_ip_i(M_{\chi}\otimes I)|\phi_i\rangle
\langle\phi_i|(M_{\chi}^{\dag}\otimes I)\big]
\!\leq\!\sum_ip_iC\big[(M_{\chi}\otimes I)|\phi_i\rangle\langle
\phi_i|(M_{\chi}^{\dag}\otimes I)\big].$ According to the Cauchy
inequality, we have $ \big|\sum_{\gamma=1}^{N_2(N_2-1)/2}C_{\alpha
\gamma}[|\chi\rangle]C_{\gamma\beta}[\rho_{\$}]\big|^2
\!\leq\!\sum_{\gamma=1}^{N_2(N_2-1)/2}\big|C_{\alpha
\gamma}[|\chi\rangle]\big|^2\sum_{\gamma'=1}^{N_2(N_2-1)/2}\big|C_{\gamma'\beta}[\rho_{\$}]\big|^2.
$ In terms of Eq. (\ref{ev4}) we get
\begin{eqnarray}
\label{ev13} C(\rho')\!&\leq&\!\sum_ip_iC\big[(M_{\chi}\otimes
I)|\phi_i\rangle\langle\phi_i|
(M_{\chi}^{\dag}\otimes I)\big]\nonumber\\
%&\leq&\frac{N_2}{2}\sqrt{\sum_{\alpha=1}^{N_1(N_1-1)/2}\quad
%\sum_{\gamma=1}^{N_2(N_2-1)/2}|C_{\alpha\gamma}[|\chi\rangle]|^2}\nonumber\\
%&&\times\sum_ip_i\sqrt{\sum_{\gamma'=1}^{N_2(N_2-1)/2}\quad
%\sum_{\beta=1}^{N_2(N_2-1)/2}|C_{\gamma'\beta}[|\phi_i\rangle]|^2}\nonumber\\
\!&\leq&\!\frac{N_2}{2}C[|\chi\rangle]\sum_ip_iC[|\phi_i\rangle]=\frac{N_2}{2}C[|\chi\rangle]C[\rho_{\$}].
\end{eqnarray}
This inequality can be generalized to the case that the initial
state $\rho_0$ is a mixed one,
\begin{eqnarray}
\label{ev14} C\big[(
I\otimes\$)\rho_0\big]\leq\frac{N_2}{2}C(\rho_0)C[\rho_{\$}].
\end{eqnarray}

If we consider bipartite states in $N_1\otimes2$ system, the result
(\ref{ev4}) can be generalized for arbitrary noisy channels $\$$. So
we get the following corollary: if the pure initial state is a
bipartite $N_1\otimes2$ one, while the second subsystem is subject
to an arbitrary noisy channel $\$$, we have the following evolution
equation of concurrence,
\begin{eqnarray}
\label{thm2} \label{ev8} C[\rho']=C[|\chi\rangle] {C}[\rho_{\$}].
\end{eqnarray}

This corollary is proved as follows. Without loss of generality, we
suppose that $\rho_{\$}$ is a mixed state. By using the procedure of
the optimal pure state decomposition adopted in Ref.
\cite{wootters}, there must exist an optimal pure state
decomposition for a $2\otimes2$ mixed state
\begin{equation}
\label{ev5}
 \rho_{\$}=\sum_ip_i|\psi_i\rangle\langle\psi_i|,
\end{equation}
 such that ${C}[\rho_{\$}]\!=\!{C}[|\psi_i\rangle]$, $\forall i$, are satisfied.
Suppose the pure decomposition Eq. (\ref{ev5}) is not optimal for
concurrence of $\rho'$ in terms of Eq. (\ref{vec0}). Then there must
exist another decomposition other than Eq. (\ref{ev5}), $
\rho_{\$}\!=\!\sum_iq_i|\psi'_i\rangle\langle\psi'_i|,$ which is an
optimal pure state decomposition of $\rho'=(M_{\chi}\otimes
I)\rho_{\$}(M_{\chi}^{\dag}\otimes I)$. In terms of Eq. (\ref{ev4}),
we have
\begin{eqnarray}
\label{ev7} C[\rho']=\sum_iq_iC[(M_{\chi}\otimes I)|\psi'_i\rangle]
%&=&C[|\chi\rangle]\sum_iq_i C[|\psi'_i\rangle]
\geq
C[|\chi\rangle]C[\rho_{\$}].
\end{eqnarray}
However, in terms of the optimal pure state decomposition
(\ref{ev5}) and convexity, we have
\begin{eqnarray}
\label{ev77} C[\rho']<\sum_ip_iC\big[(M_{\chi}\otimes
I)|\psi_i\rangle\big]
%&=&C[|\chi\rangle]\sum_ip_iC[|\psi_i\rangle]
=C[|\chi\rangle]C[\rho_{\$}].
\end{eqnarray} It contradicts with Eq. (\ref{ev7}).
Therefore, the optimal pure state decomposition  (\ref{ev5}) is also
optimal for concurrence (\ref{vec0}) of $\rho'$. Therefore, we get
Eq. (\ref{thm2}). \hspace{78pt}$\blacksquare$

The result (\ref{ev8}) can also be generalized to the case that the
initial state $\rho_0$ is mixed. Let
$\rho_0=\sum_ip_i|\psi_i\rangle\langle\psi_i|$ be an optimal pure
state decomposition, in the sense that the average Frobenius norm of
the concurrence matrix over this pure state decomposition is
minimal. According to convexity, we have $C[(
I\otimes\$)\rho_0]\!=\!C[\sum_ip_i(
I\otimes\$)|\psi_i\rangle\langle\psi_i|] \!\leq\!\sum_ip_iC[(
I\otimes\$)|\psi_i\rangle\langle\psi_i|].$ Using Eq. (\ref{ev8}), we
have
\begin{eqnarray}
\label{ev99} C[( I\otimes\$)\rho_0] \leq C[\rho_0]C[\rho_{\$}].
\end{eqnarray}
When $N_1=2$, the result (\ref{ev8}) reduces to the main result of
Thomas Konrad  \emph{et al}. \cite{thomas}.
%\begin{eqnarray}
%\label{ev9} {C}[(\$_{1}\otimes\$_{2})\chi] \leq
%{C}[(\$_{1}\otimes\$_{2})|\phi\rangle\langle\phi|]C[|\chi\rangle].
%\end{eqnarray}
The results (\ref{ev4}), (\ref{ev13})-(\ref{ev8}), and (\ref{ev99}))
show that the dynamics of entanglement for bipartite systems under a
one-sided noisy channel is determined by the channel's action on the
maximally entangled state.

 Let us study in which case the evolution equation of entanglement
holds for $2\otimes2$ pure initial state under the influence of
local two-sided channel $\$_{1}\otimes\$_{2}$. The channels $\$$ can
usually be expressed as Kraus operators \cite{kraus}. First consider
the phase noise channel $\$_{1}$ satisfying
$\mathrm{Tr}(\sigma_x\$_{1})=\mathrm{Tr}(\sigma_y\$_{1})=0,$ we have
$ {C}[(\$_{1}\otimes\$_{2})\chi] =
{C}[(\$_{1}\otimes\$_{2})|\phi\rangle\langle\phi|]C(|\chi\rangle) $
for any initial state $|\chi\rangle=a|00\rangle+b|11\rangle$ in
Schmidt expression. If $\$_{1}$ satisfies the condition
$\mathrm{Tr}(\sigma_z\$_{1})=\mathrm{Tr}(\$_{1})=0$, for
$|\chi\rangle=a|00\rangle+b|11\rangle$, we have $
{C}[(\$_{1}\otimes\$_{2})\chi]=
{C}[(\$_{1}\otimes\$_{2})|\phi\rangle\langle\phi|]C(|\chi\rangle)$;
 for $|\chi\rangle=a|01\rangle+b|10\rangle$, we have
$ {C}[(\$_{1}\otimes\$_{2})\chi]= {C}[(
I\otimes\$_{2}\$_{1})|\phi\rangle\langle\phi|]C(|\chi\rangle). $

For the case that the initial state is mixed, let us consider
\cite{echain} $\rho_0= \tiny{\left(
   \begin{array}{cccc}
   a & 0 & 0 & 0\\
   0 & b & d & 0\\
   0 & d^*& c& 0\\
   0 & 0 & 0 & 0
   \end{array}
\right )}. $ For phase noise channel $\$_p$, we still have ${C}[(
I\otimes\$_p)\rho_0]= {C}[\rho_0]C[\rho_{\$_p}]$.

\emph{Application to two realistic systems}---Let us consider a
three-qubit system with the third qubit exposed to a phase noise
channel $\$_p$. The phase noise channel $\$_p$ can be expressed as
Kraus operators: $ K_1\!=\!\tiny{\left(
   \begin{array}{cc}
   \nu& 0 \\
   0 & 1
   \end{array}
\right ), K_2\!=\!\left(
   \begin{array}{cc}
   \omega& 0 \\
   0 & 0
   \end{array}
\right )} $, where the time-dependent Kraus matrix elements are
$\nu\!=\!\mathrm{exp}[-\Gamma t]$ and $\omega\!=\!\sqrt{1-\nu^2}$.
 We study how the residual
entanglement \cite{coffman,yu} evolves. For the initial state
$|\chi\rangle_{ABC}\!=\!\alpha|001\rangle\!+\!\beta|010\rangle\!+\!\gamma|100\rangle$,
generalized W state, it follows that $\mathrm{Tr}_A\big[( I\otimes
I\otimes\$_p)|\chi\rangle_{ABC}\langle\chi|\big]\!=\! (
I\otimes\$_p)\mathrm{Tr}_A\big[|\chi\rangle_{ABC}\langle\chi|\big]\!=\!(
I\otimes\$_p)\rho_{BC}$, since the partial trace is a local
operation. Therefore, we obtain
$\rho_{BC}\!=\!\mathrm{Tr}_A|\chi\rangle_{ABC}\langle\chi|\!=\!|\gamma|^2|00\rangle\langle00|
+(\alpha|01\rangle\!+\!\beta|10\rangle)(\langle01|\alpha^*\!+\!\langle10|\beta^*)$
and
$\rho_{AC}\!=\!\mathrm{Tr}_B|\chi\rangle_{ABC}\langle\chi|\!=\!|\beta|^2|00\rangle\langle00|
+(\alpha|01\rangle\!+\!\gamma|10\rangle)(\langle01|\alpha^*\!+\!\langle10|\gamma^*)$.
Due to the fact that $\rho_{AB}$ and $\rho_{AC}$ are special cases
of $\rho_0$ in \cite{echain}, we immediately get ${C}[(
I\otimes\$_p)\rho_{BC}]\!=\!C[\rho_{BC}]
{C}[\rho_{\$_p}]\!=\!2|\alpha\beta|\mathrm{exp}[-\Gamma t]$ and
${C}[( I\otimes\$_p)\rho_{AC}]\!=\!
{C}[\rho_{AC}]C[\rho_{\$_p}]\!=\!2|\alpha\gamma|\mathrm{exp}[-\Gamma
t]$. If we regard $AB$ subsystem as a whole 4 dimensional system, we
can calculate $C\big[( I\otimes I\otimes\$_p)\rho_{AB:C}\big] $ by
means of Eq. (\ref{ev8}), $ C\big[( I\otimes
I\otimes\$_p)\rho_{AB:C}\big]
%\!=\!C\big[( I_{AB}\otimes\$_p)\rho_{AB:C}\big]
\!=\!C\big[\rho_{AB:C}\big]C[\rho_{\$_p}]
\!=\!2|\alpha|\sqrt{|\gamma|^2\!+\!|\beta|^2}\mathrm{exp}[-\Gamma
t]$. Therefore the residual entanglement
$\tau_{C(AB)}\!=\!C\big[\rho'_{AB:C}\big]^2\!-\!{C}[\rho'_{BC}]^2
\!-\!{C}[\rho'_{AC}]^2\!=\!0$. This indicates that the residual
entanglement $\tau_{C(AB)}$ of this initial state keeps invariant
when the third subsystem is exposed to a dephasing channel $\$_p$.

\begin{figure}[!b]
\begin{center}
\scalebox{0.56}{\includegraphics*[135pt,10pt][437pt,215pt]{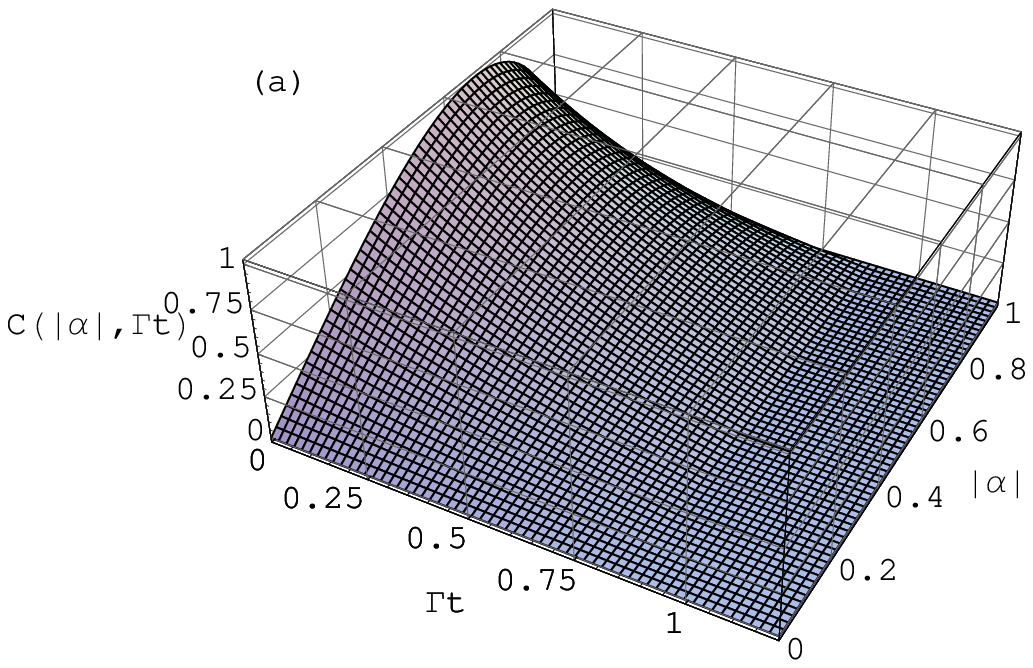}}
\scalebox{0.51}{\includegraphics*[52pt,0pt][350pt,215pt]{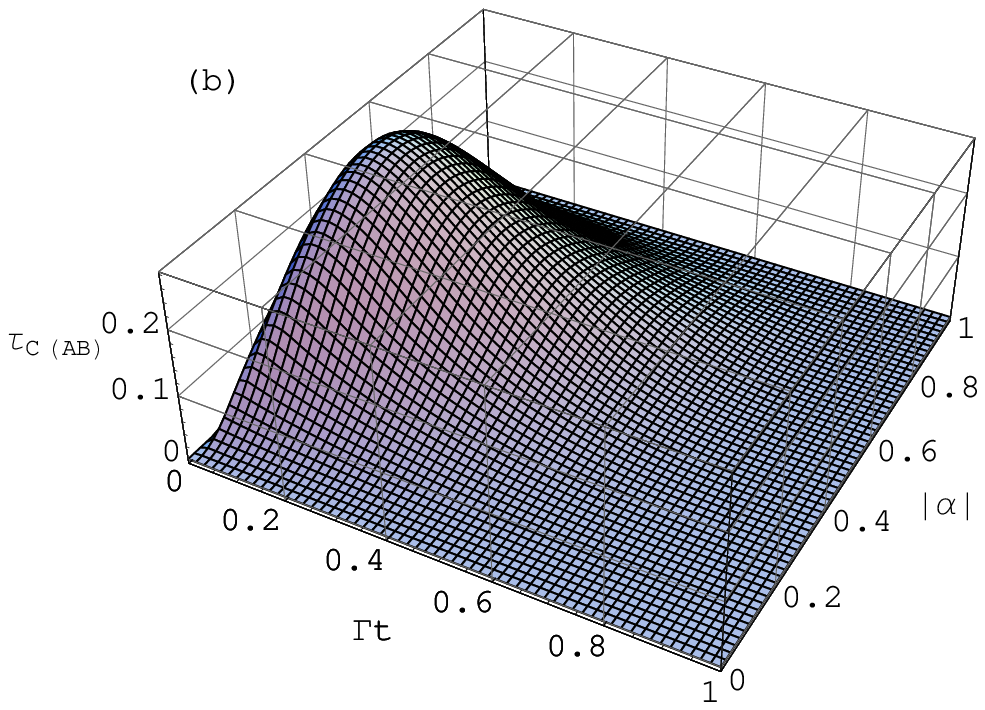}}
\caption{(a) The concurrence $C_{AB:C}(|\alpha|,t)$ vs $\Gamma t$
and amplitude $|\alpha|$, whree $\Gamma$ is generalized amplitude
decay rate. The system of AB and of C disentangle completely and
abruptly in just a finite time for all $|\alpha|$ in the range
shown. (b) Dependence of the residual entanglement $\tau_{C(AB)}$ on
$\Gamma t$ and amplitude $|\alpha|$ for $|\beta|=|\gamma|$. The
residual entanglement gets maximized value at the point $\Gamma
t=0.0936$ and $|\alpha|=0.4996$.}
\end{center}
\end{figure}
Next we consider the \textit{generalized amplitude damping} channel,
$\$_{GAD}$, describing the effect of dissipation to an environment
at finite temperature, which is the case relevant to NMR quantum
computation. The channel usually adopts the form as: $
K_0=\sqrt{p}\tiny{\left(
   \begin{array}{cc}
   1& 0 \\
   0 &\nu
   \end{array}\right )},
K_1=\sqrt{p}\tiny{\left(
   \begin{array}{cc}
   0& \omega \\
   0 & 0
   \end{array}
\right )}, K_2=\sqrt{1-p}\tiny{\left(
   \begin{array}{cc}
   \nu& 0 \\
   0 & 1
   \end{array}
\right )}, K_3=\sqrt{1-p}\tiny{\left(
   \begin{array}{cc}
   0& 0 \\
   \omega & 0
   \end{array}
\right ) } $.  Without loss of generality, set $p=\frac{1}{2}$. In
terms of Eq. (\ref{ev8}), we obtain
$C_{AB:C}\big(|\alpha|,t\big)=C\big[( I\otimes
I\otimes\$_{GAD})\rho_{AB:C}\big]=|\alpha|\sqrt{1-|\alpha|^2}
(\mathrm{exp}[-2\Gamma t]+2\mathrm{exp}[-\Gamma t]-1)$. As shown in
Fig. 1 (a), over a continuous range of $|\alpha|$ values,
$C_{AB:C}\big(|\alpha|,t\big)$ actually goes abruptly to zero in a
finite time and remains zero thereafter. This is  the ``entanglement
sudden death" (ESD) effect \cite{karol,yuting1,mp}. The residual
entanglement is calculated as
$\tau_{C(AB)}=C^2_{AB:C}-C^2_{AB}-C^2_{AC}$. For simplicity, we
focus on $|\beta|=|\gamma|$. It is illustrated in Fig. 1 (b) that
the residual entanglement terminates completely after a finite
interval and remains zero thereafter, and it reaches a maximal value
at the point $\Gamma t=0.0936$ and $|\alpha|=0.4996$.

The evolution of entanglement is basically related to the
Hamiltonian of a physical system. For example, consider the ground
state of an NMR system,
$|\psi\rangle\!=\!\sqrt{1/3}|02\rangle\!-\sqrt{1/3}|11\rangle\!+\!\sqrt{1/3}|20\rangle$,
in which the coupling Hamiltonian of two spin$\!-\!1$ nuclei can be
expressed as: $\mathcal {H}\!=\!J\hat{S}_1\hat{S}_2$, $J\!>\!0$,
where $\hat{S}_1$ and $\hat{S}_2$ are spin operators of nucleus 1
and 2 respectively. Supposing that $|\psi\rangle$ is the initial
state, under the single-sided relaxation operation
$M\!=\!diag(e^{-\Gamma_2t},e^{-\Gamma_1t},1)$,
$\Gamma_2\geq\Gamma_1\!>\!0$ \cite{note}, the final state becomes
$\rho'\!=\!( I\otimes M)|\psi\rangle\langle\psi|( I\otimes
M^{\dagger})/p$, where $p\!=\!Tr\big[( I\otimes
M)|\psi\rangle\langle\psi|( I\otimes M^{\dagger})\big]$. In terms of
Eq. (\ref{ev4}), we obtain the dependence of entanglement on t,
$c\big[\rho'\big]\!=\!\sqrt{4\big[(e^{-2\Gamma_1t}\!+\!1)(e^{-2\Gamma_2t}\!+\!1)\!-\!1\big]}
/\big[1\!+\!e^{-2\Gamma_1t}\!+\!e^{-2\Gamma_2t}\big]$. When
$t\gg\frac{1}{\Gamma_1}$, $\rho'\rightarrow|20\rangle$.

\emph{Conclusions}.---In summary, we have investigated the time
evolution of entanglement for arbitrary bipartite systems, with one
part subject to interactions with environments. Explicit expressions
are derived for bipartite systems and a general factorization law is
obtained for multi-qubit or qubit-qudit systems with one qubit
undergoing the action of a noisy channel. It allows one to know the
time evolution of entanglement for an arbitrary initial state, if
one knows the time evolution of entanglement for the bipartite
maximally entangled state. The later only depends on the detailed
noisy channel and has nothing to do with the initial states.  Our
results can be used to infer the evolution of entanglement under
certain time-continuous influences of the environment. Due to the
fact that all the quantities of entanglement measure can be
evaluated efficiently for pure states, it can help the experimental
characterization of entanglement dynamics. Moreover, the results can
be also directly applied to input/output processes such as gates
used in sequential quantum computing. As applications we have
studied the entanglement evolution of the generalized three-qubit W
state, with one qubit undergoing the action of phase noise channel
and generalized amplitude damping channel respectively. We also
obtain the evolution of entanglement for the ground state in an  NMR
system with one subsystem subject to a decay channel.

This work is supported by NSFC under grant 90406017, 60525417,
10675086, 10740420252, and the NKBRSFC under grant 2004CB318000,
2005CB724508 and 2006CB921400.


\begin{thebibliography}{99}
\bibitem{bennett} C. H. Bennett and D. P. DiVincenzo, Nature (London) \textbf{404}, 247
(2000).
\bibitem{nielsen} M. A. Nielsen and I. L. Chuang, \emph{Quantum Computation and Quantum Information}
(Cambridge University Press, Cambridge, 2000).
\bibitem{thomas} T. Konrad, F. De Melo, M. Tiersch, C. Kasztelan, A.
Arag\~{a}o, and A. Buchleitner,  Nature Physics \textbf{4}, 99
(2008).
\bibitem{markus} M. Tiersch, F. De Melo, and A. Buchleitner,
arXiv:0804.0208v1.
\bibitem{yuting} T. Yu and J. H. Eberly, Phys. Rev. Lett. \textbf{97}, 140403
(2006).
\bibitem{dodd}P. J. Dodd and J. J. Halliwell, Phys. Rev. A \textbf{69}, 052105
(2004).
\bibitem{karol} K. \.{Z}yczkowski, P. Horodecki, M. Horodecki, and R.
Horodecki, Phys. Rev. A \textbf{65}, 012101 (2001).
\bibitem{frank} F. Verstraete, J. Dehaene, and B. DeMoor, Phys. Rev. A \textbf{64}, 010101(R)
(2001).
%\bibitem{gour1} G. Gour, Phys. Rev. A 71, 012318 (2005);
%G. Gour, Phys. Rev. A 72, 042318 (2005).
\bibitem{cluster}
K. Chen, C.M. Li, Q. Zhang, Y.A. Chen, A. Goebel, S. Chen, A. Mair
and J.W. Pan, Phys. Rev. Lett. \textbf{99}, 120503 (2007); Y.
Tokunaga, S. Kuwashiro, T. Yamamoto, M. Koashi and N. Imoto, Phys.
Rev. Lett. \textbf{100}, 210501 (2008).
\bibitem{Pan}
Q. Zhang, A. Goebel, C. Wagenknecht, Y.A. Chen, B. Zhao, T. Yang, A.
Mair, J. Schmiedmayer and J.W. Pan Nature Phys. \textbf{2}, 678-682
(2006).
\bibitem{crypt} T. Durt, N. J. Cerf, N. Gisin, and M. \.{Z}ukowski, Phys. Rev. A \textbf{67}, 012311 (2003);
S. P. Walborn, D. S. Lemelle, M. P. Almeida, and P. H. Souto
Ribeiro, Phys. Rev. Lett. \textbf{96}, 090501 (2006).
\bibitem{w} W. D\"{u}r, G. Vidal, and J. I. Cirac, Phys. Rev. A \textbf{62}, 062314 (2000).
\bibitem{rungta} P. Rungta, V. Bu\v{z}ek, C. M. Caves, M. Hillery, and G. J. Milburn, Phys. Rev. A \textbf{64},
042315 (2001).
\bibitem{wootters} W. K. Wootters, Phys. Rev. Lett. \textbf{80}, 2245 (1998).
\bibitem{measureable} F. Mintert, M. Ku\'{s}, and A. Buchleitner, Phys. Rev. Lett.
\textbf{95}, 260502 (2005); S. P. Walborn, P. H. Souto Ribeiro, L.
Davidovich, F. Mintert, and A. Buchleitner, Nature (London)
\textbf{440}, 1022 (2006).  The quantity of $|C_{\alpha\beta}|^2$
can be obtained by performing the single joint measurement on the
two copies of the pure state of interest. Actually it is
proportional to the probability of observing the two copies of the
first subsystem in an antisymmetric state, which is proposed in the
above articles. The quantity of concurrence C can also be obtained
in the similar way.
\bibitem{wootters2} W. K. Wootters, Quantum Inf. Comput. \textbf{1}, 27 (2001).
\bibitem{v1} K. Audenaert, F. Verstraete and B. De Moor, Phys. Rev. A \textbf{64}, 052304 (2001); S. J. Akhtarshenas, J. Phys. A \textbf{38}, 6777 (2005).
\bibitem{albev} S. Albeverio and S. M. Fei, J. Opt. B \textbf{3}, 223 (2001).
\bibitem{gisin} N. Gisin, Phys. Lett. A \textbf{210}, 151 (1996).
\bibitem{kraus} K. Kraus, \emph{States, Effect, and Operations:
Fundamental Notions in Quantum Theory} (Springer-Verlag, Berlin,
1983).
\bibitem{echain} W. K. Wootters, Contemp. Math. \textbf{305}, 299 (2002).
\bibitem{coffman} V. Coffman, J. Kundu, and W. K. Wootters, Phys. Rev. A  \textbf{61}, 052306
(2000).
\bibitem{yu} C. S. Yu and H. S. Song, Phys. Rev. A  \textbf{71}, 042331
(2005).
\bibitem{yuting1} T. Yu and J. H. Eberly, Opt. Commun. \textbf{264}, 393
(2006).
\bibitem{mp} M. P. Almeida, F. de Melo, M. Hor-Meyll, A. Salles, S. P. Walborn,
P. H. Souto Ribeiro, and L. Davidovich, Science \textbf{316}, 579
(2007).
%\bibitem{nmr1} R. R. Ernst, Principles of Nuclear Magnetic Resonance in One and Two
%Dimensions, International Series of Monographs on Chemistry (Oxford
%University Press, 1990).

%\bibitem{nmr2} M. H. Levitt, Spin Dynamics (Wiley, 2001).
\bibitem{note} This single-sided relaxation operation can be realized by
performing a measurement (e.g. a projective measurement) on the
system of interest and the environment.
\end{thebibliography}
\end{document}